\documentclass[prl,reprint,twocolumn,amsmath,amssymb,showpacs]{revtex4}
\usepackage{graphicx}
\usepackage{dcolumn}
\usepackage{bm}

\usepackage{color}
\definecolor{verde}{rgb}{0,0.5,0}
\usepackage{psfrag}
\graphicspath{{fig_rev/}{nuovo/}{Re150/}{nov2008/}}
\input{colornam.sty}
\begin{document}

\preprint{Under submission to PRL}

\title{Small scale anisotropy in turbulent shearless mixing}
\author{Daniela Tordella}%
 \email{daniela.tordella@polito.it}
\affiliation{%
Dipartimento di Ingegneria
Aeronautica e Spaziale, Politecnico di Torino,10129 Torino, Italy\\
International Center for Turbulence Cooperation, ICTR.}%
\author{Michele Iovieno}
\affiliation{%
Dipartimento di Ingegneria
Aeronautica e Spaziale, Politecnico di Torino,10129 Torino, Italy\\
International Center for Turbulence Cooperation, ICTR.}%
\date{\today}

\begin{abstract}
The generation of small-scale anisotropy in turbulent shearless mixing is numerically investigated. Data from direct numerical simulations at Taylor Reynolds' numbers between 45 and 150 show that there is not only a significant departure of the longitudinal velocity derivative moments from the values found in homogeneous and isotropic turbulence, but that the variation of skewness has an opposite sign for the components across the mixing layer and parallel to it. The anisotropy induced by the presence of a kinetic energy gradient has a very different pattern from the one generated by an homogeneous shear.
The transversal derivative moments in the mixing are in fact found to be very small, which highlights that smallness of the transversal moments is not a sufficient condition for isotropy.
\end{abstract}

\pacs{47.27+, 47.51+ }
\maketitle

Laboratory and numerical results are continuously being generated on the small-scale features of turbulence dynamics (see e.g. \cite{sa97,ws00,ws02, ssy03,syb03, ws04,ob06,tib08}). Turbulent flows  contain a wide range of scales and each range is characterized by its own physics. For example, energy dissipation takes place at small-scales. However, the process is linked to the large scales of the system and the existence of a long-range interaction should not be excluded. 

A manifestation of the non-universal behavior of small-scales is closely related to small-scale anisotropy, which can be represented in terms  of velocity derivative statistics. In this paper, we have analyzed the velocity derivative statistics of shearless turbulent mixing in temporal decay, a very mild instance of inhomogeneity where near-isotropy could reasonably be expected. In fact, the turbulent shearless mixing layer is possibly the simplest instance of an inhomogeneous turbulent flow, because it is generated by the interaction of two isotropic turbulences in the absence of a mean shear flow. Therefore, in this flow, there is no production of turbulent kinetic energy and no mean convection. In general, this mixing shows that the behavior of statistical turbulent quantities are influenced by the presence of a kinetic energy gradient.

The shearless mixing  is a flow where a significant level of anisotropy is observed at large and small scales. The anisotropy persists at the moderate Reynolds numbers that have been reached in the numerical simulations \cite{bfkm96,kdc04,ti06,tib08,km08} and at the moderate/high numbers that have been reached in the laboratory \cite{vw89,km08}.
These studies show that the one-point velocity statistics exhibit high intermittency in the velocity component along the mixing, as indicated by the large maxima of skewness and kurtosis, and  only a mild anisotropy of the second order moments. The level of intermittency is a function of both the energy and integral scale gradients but a kinetic energy variation alone is sufficient for the onset of intermittency \cite{tib08}.

\textcolor{black}{It should be noted, that in this flow the turbulence structure is different from the homogeneous sheared turbulence \cite{sa97,ws00,ws02, ssy03, ws04}, but also from the turbulence near the fluctuating interfaces at the outer edges of turbulent shear layers, with and without free-stream turbulence \cite{bhr02,hew06,st10}. Here, the turbulent energy gradient is imposed and cannot be intensified by the continuous fluctuating interfaces produced by the instability of the mean shear. However, inside the shear-free mixing a front of high intermittency is produced and is displaced towards the low energy side of the flow \cite{ti06,tib08,vw89}. The entrainment process is active and is carried out at the level of both the large and the small fluctuations.}

Only one inhomogeneous direction is present in this flow configuration. The correlations are axisymmetric and only the longitudinal derivative moments are significantly different from zero. This kind of anisotropy is an intermediate situation between isotropic turbulence (where only longitudinal moments are present and equal, see, for example, \cite{sa97, is07, bnt08}), and homogeneous sheared turbulence, where both longitudinal and transversal moments are generated, see \cite{ws00, ws02}.
In the last case the literature shows that, as the Reynolds number is increased, the longitudinal derivative moment in the direction of the mean flow increases  while the transversal odd moments (of order 3,5 and 7) are small in comparison to the even moments. It should be noticed that the information on the other two longitudinal derivative moments (in a Cartesian reference frame, those in the plane orthogonal to the mean flow) is not available at the moment.


\begin{figure}
\centering
\includegraphics[width=0.90\columnwidth]{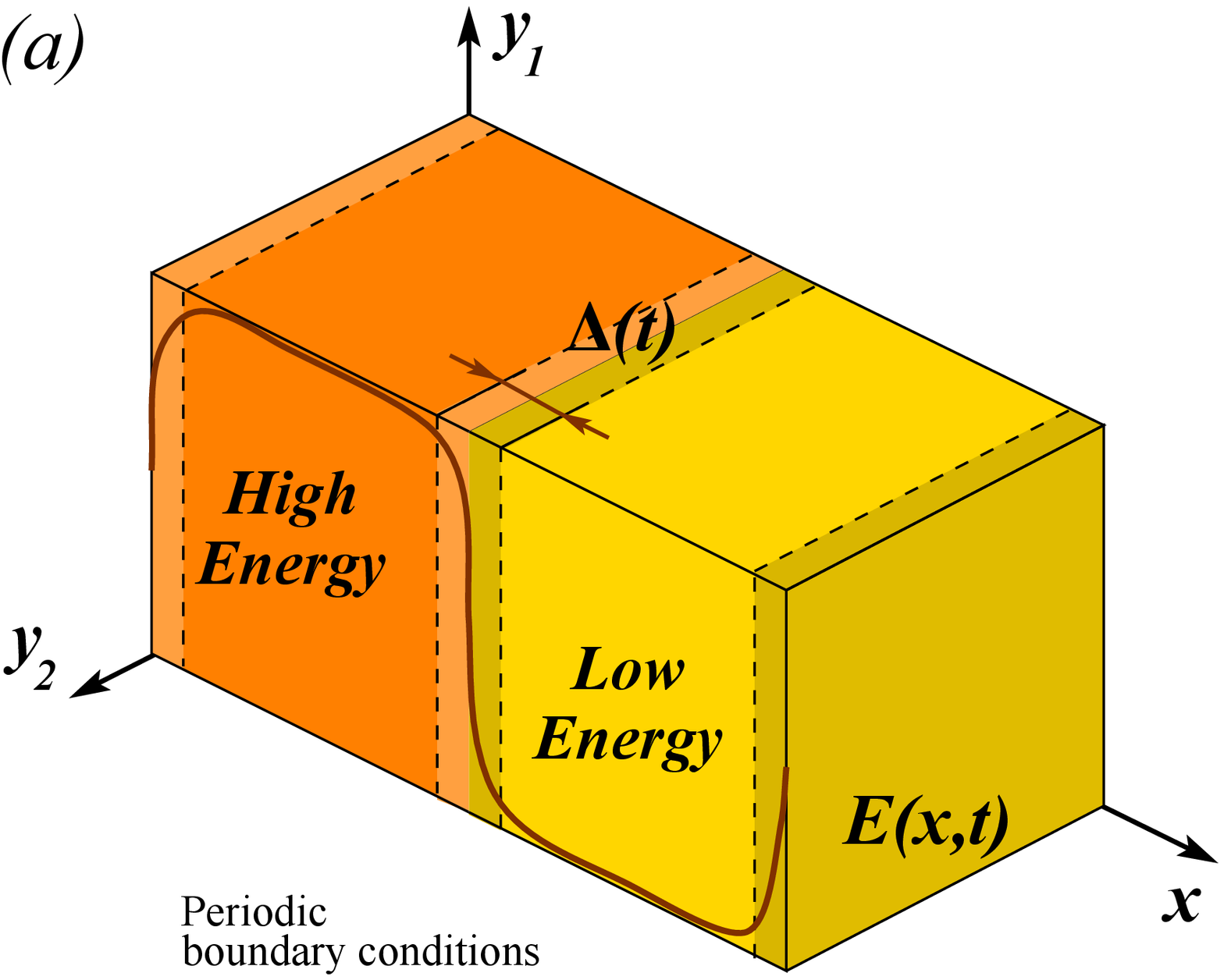}\\[0.1mm]
\includegraphics[width=0.90\columnwidth]{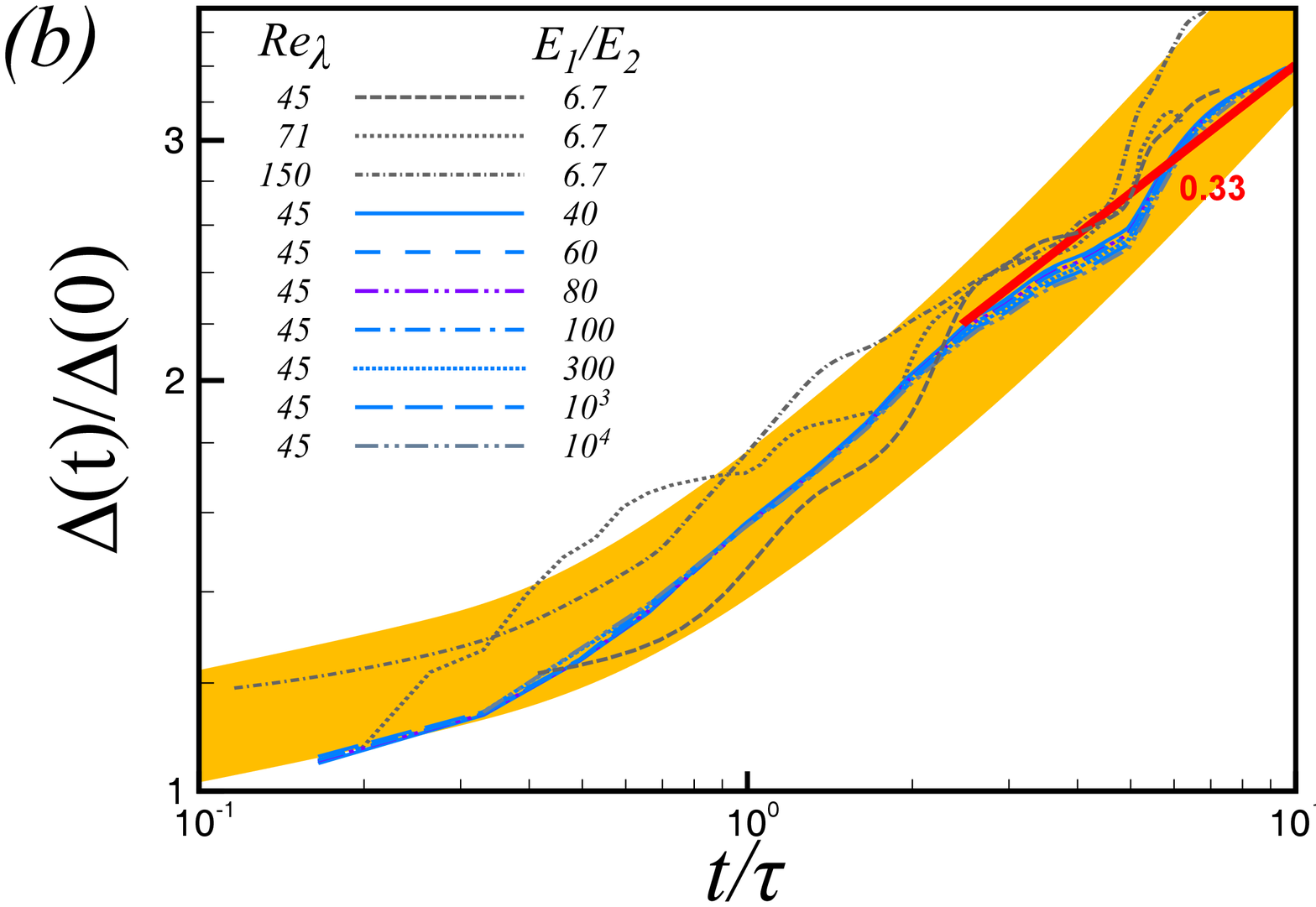}
\caption{Turbulent shearless mixing layer: (a) Scheme of the computational domain and boundary conditions. (b) Time evolution of the mixing layer thickness, varying the initial energy ratio $E_1/E_2$. The mixing layer thickness $\Delta(t)$ is conventionally defined as the distance between the points with normalized energy values $(E-E_1)/(E_1-E_2)$ equal to 0.25 and 0.75. The color band is the power law fitting with exponent 0.46. Exponent 0.33, which is indicated in the figure, is the value measured after the initial transient of the simulations has elapsed.}
\label{fig.delta}
\end{figure}

\begin{figure*}
\centering
\psfrag{S}{\hspace*{-9mm}\large\textcolor{red}{$S_{\partial u/\partial x}$,} \textcolor{verde}{$S_{\partial v/\partial y}$}}
\psfrag{K}{\hspace*{-8mm}\large\textcolor{red}{$K_{\partial u/\partial x}$,} \textcolor{verde}{$K_{\partial v/\partial y}$}}
\psfrag{z}{\hspace*{-2mm}\large\textcolor{black}{\small$\eta=x/\Delta(t)$}}
\psfrag{y}{\small $\partial u/\partial x$}
\psfrag{x}{\small $\partial v/\partial y$}
\includegraphics[width=0.46\textwidth]{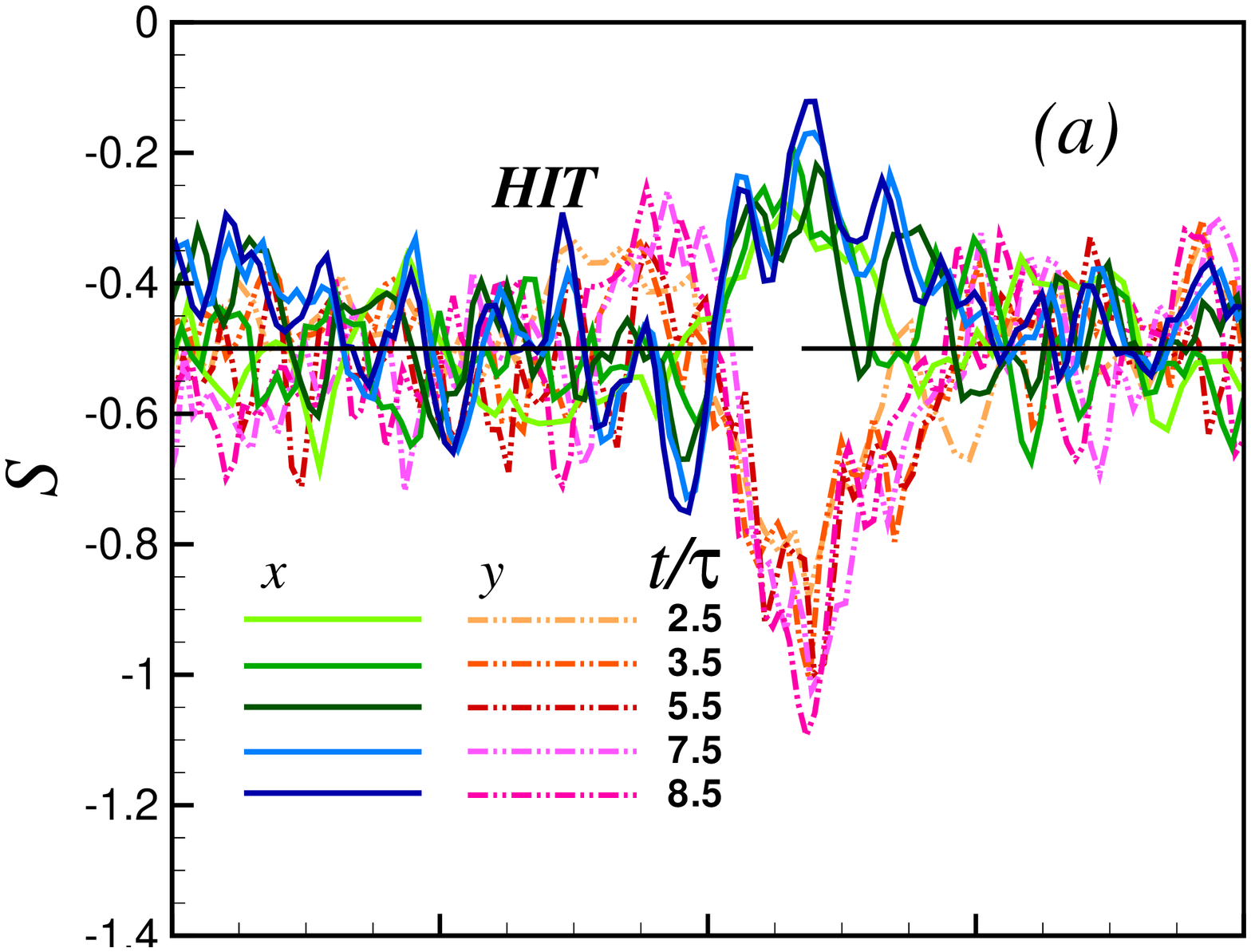} 
\includegraphics[width=0.46\textwidth]{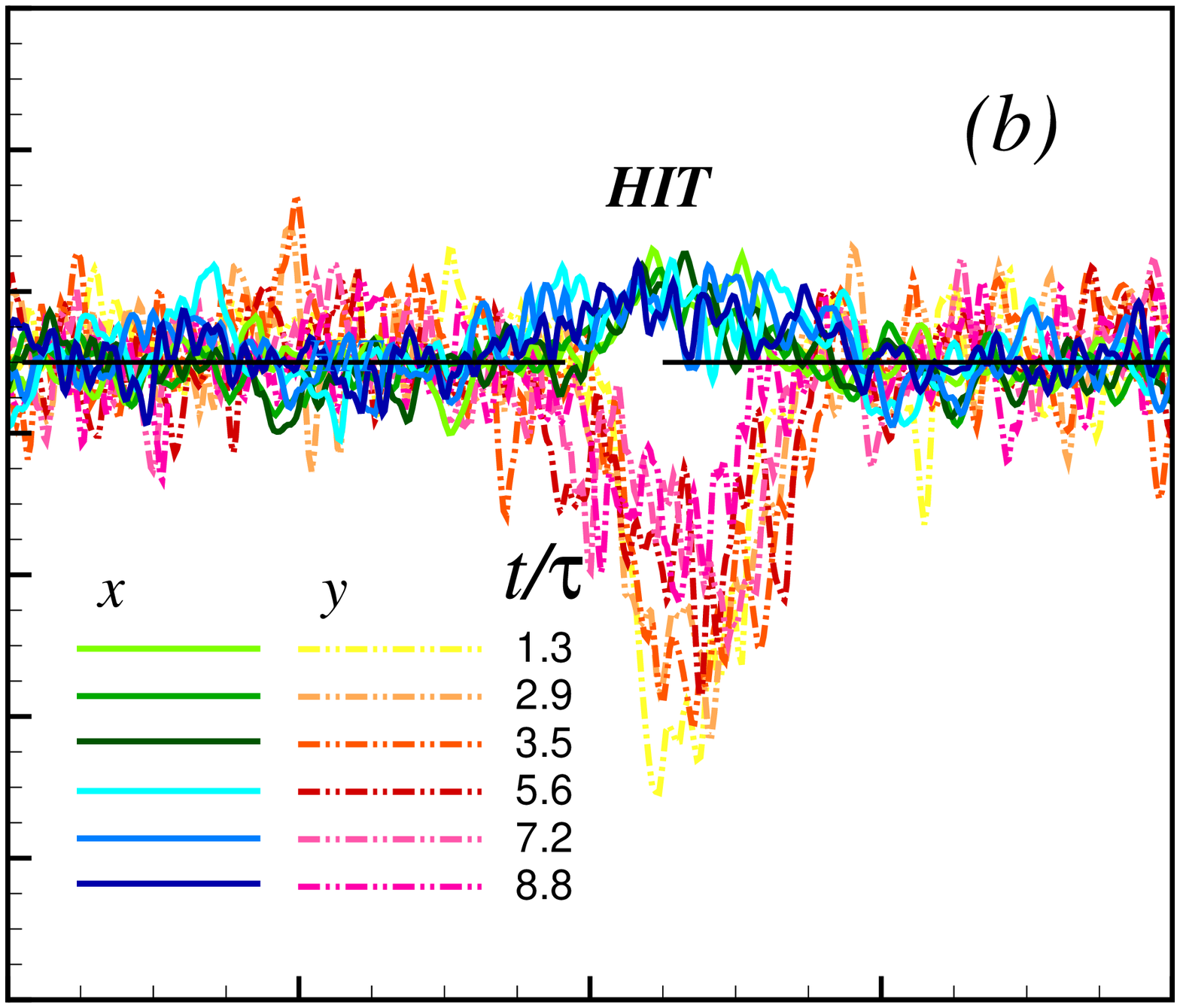}\\ 
\includegraphics[width=0.46\textwidth]{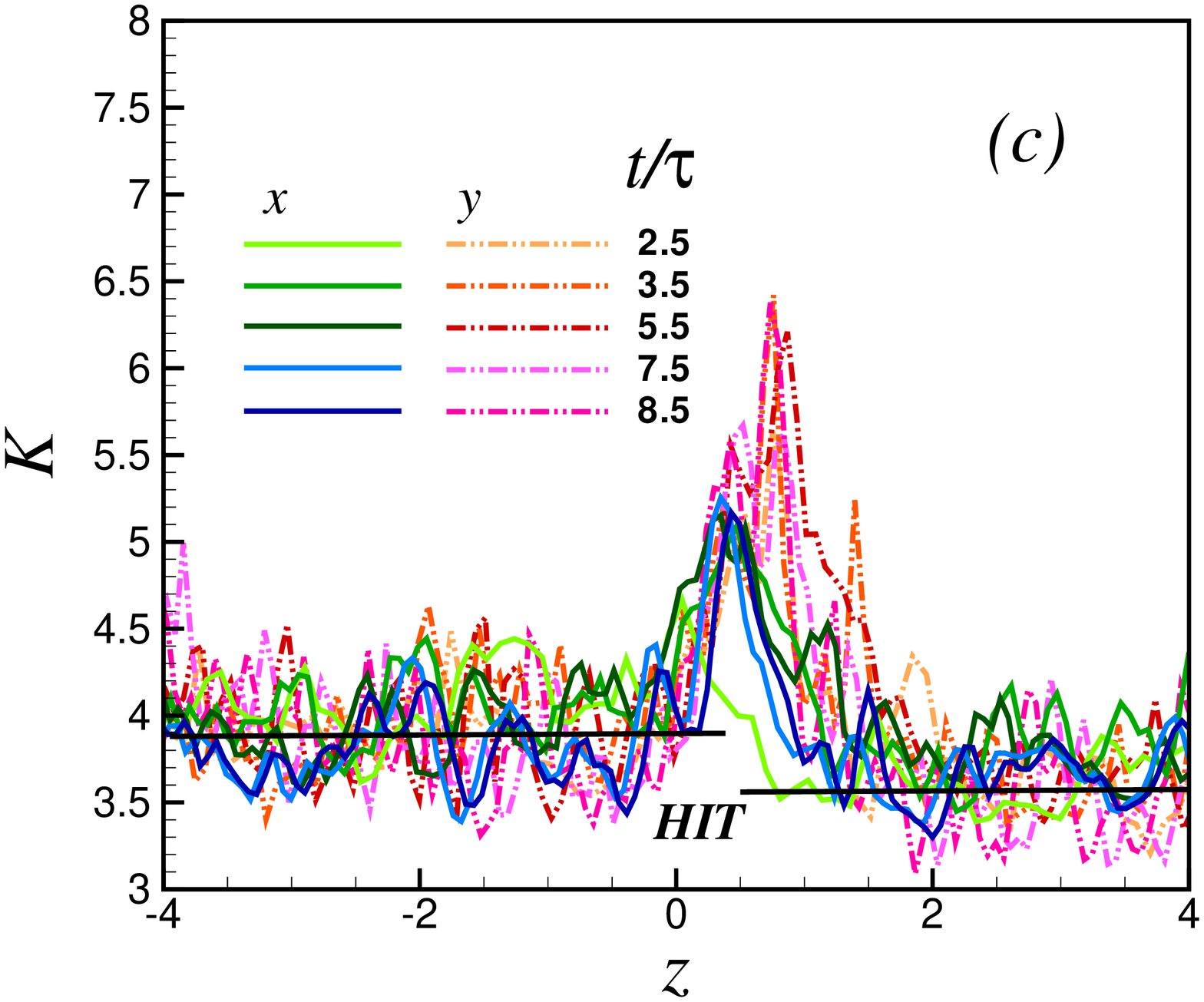} 
\includegraphics[width=0.46\textwidth]{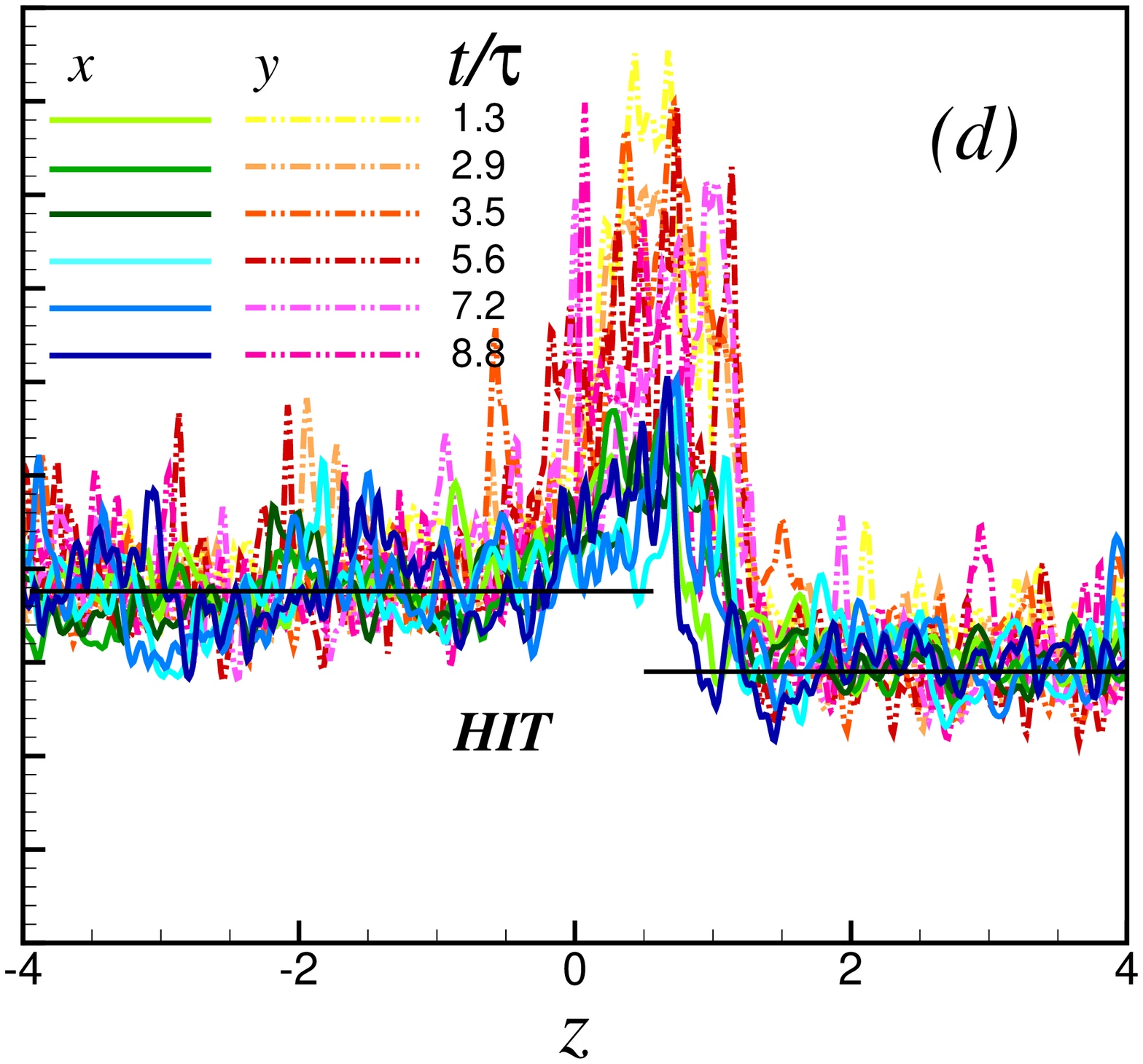} 
\caption{Longitudinal derivative skewness and kurtosis distributions: the dash-dot lines represent the statistics of $\partial u/\partial x$, the continuous lines represent the statistics of $\partial v/\partial y$. The high energy region has a Taylor microscale Reynolds number equal to 45 in parts (a) and (c) and 150 in parts (b) and (d). The thick horizontal lines represent the values of the longitudinal skewness and kurtosis in the isotropic flow regions.}
\label{fig.sk}
\end{figure*}



\begin{figure}
\centering
\includegraphics[width=\columnwidth]{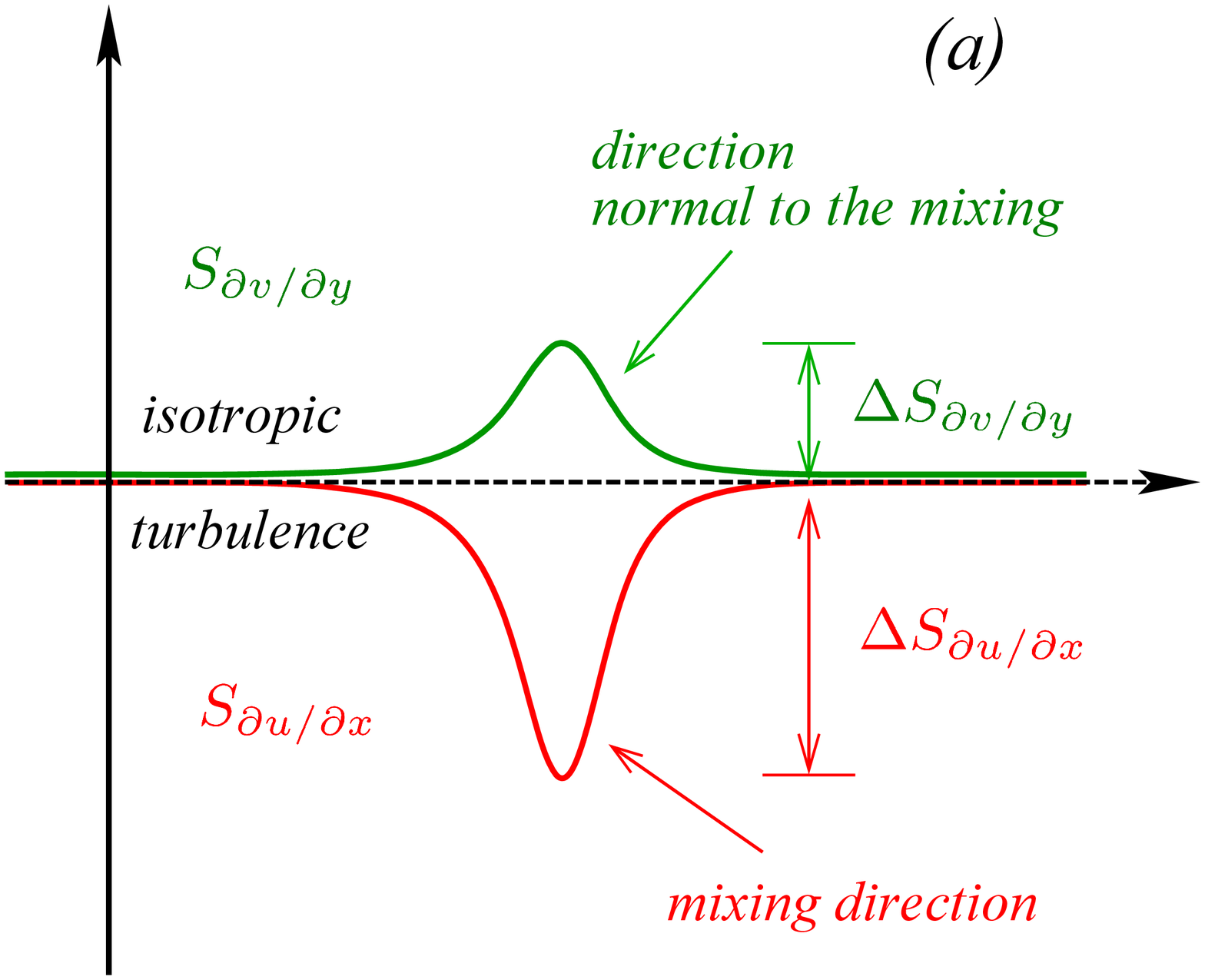}\\

\includegraphics[width=\columnwidth]{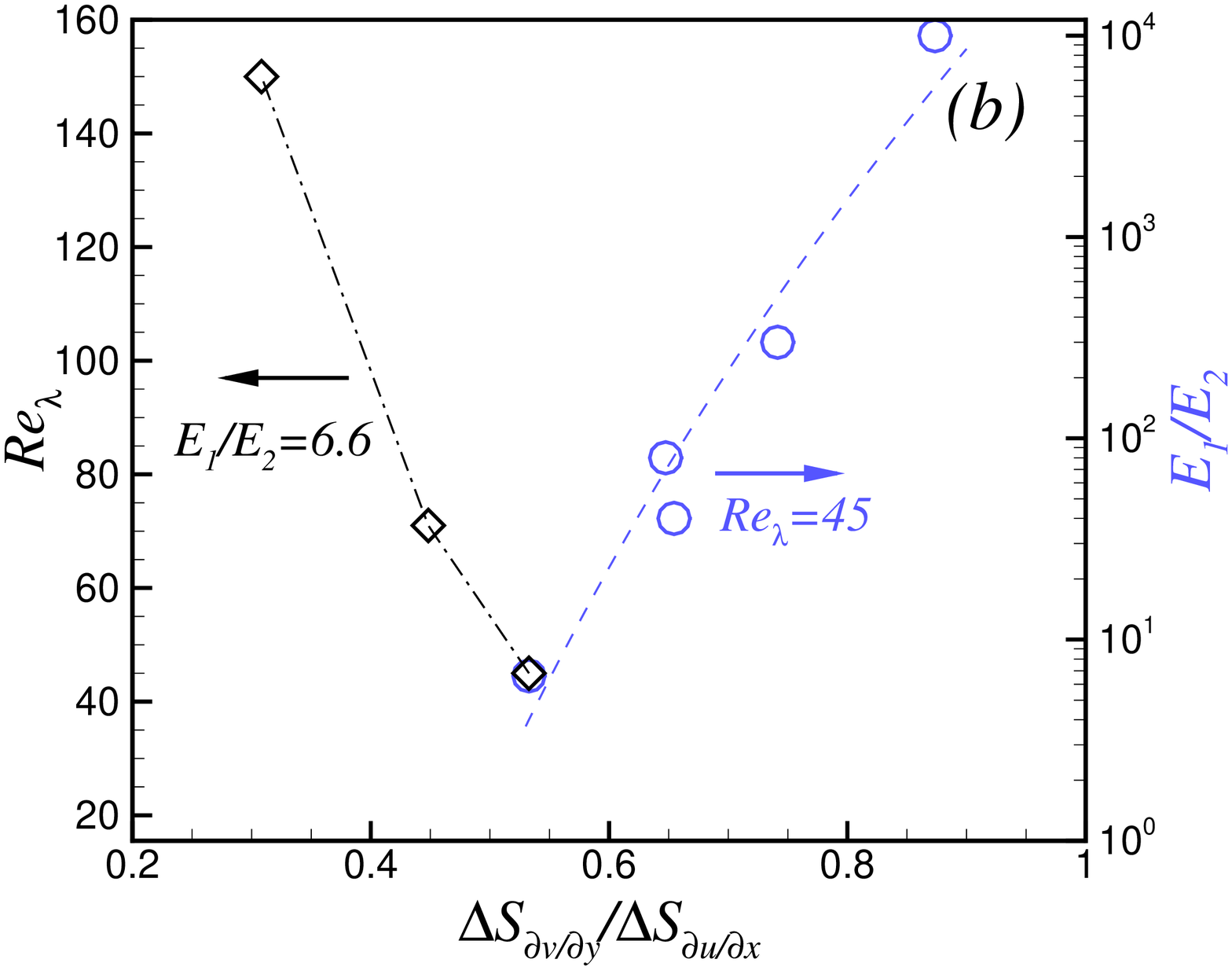}
\caption{(a) Scheme of the general behaviour of the longitudinal derivative skewness in the shearless mixing layer. (b) Anisotropy of the longitudinal derivative statistics variations inside the mixing layer as a function of the energy ratio at $Re_\lambda=45$ (circles) and as a function of the Reynolds number at $E_1/E_2=6.6$ (squares). All the quantities have been computed in the centre of the mixing layer, $\Delta S$ is the modulus of the difference between the values of the velocity derivative skewness in the center of the layer and in isotropic condition.}
\label{fig.schema_long}
\end{figure}

\begin{figure}
\includegraphics[width=\columnwidth]{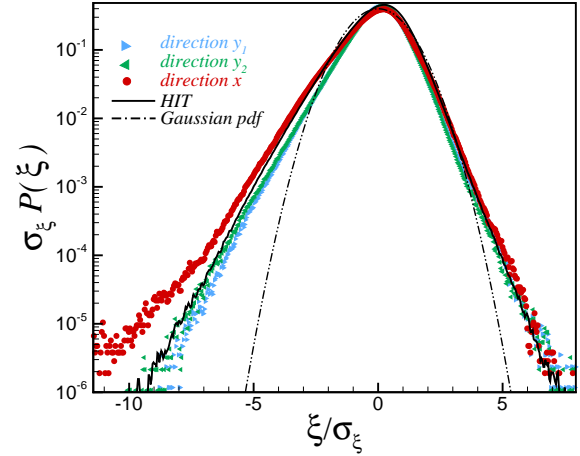}
\caption{Normalized probability density function of the longitudinal derivatives; data at $t/\tau=3.5$ in the centre of the mixing layer from the simulation at $Re_\lambda=150$, $E_1/E_2=6.6$; $\xi={\partial u_i/\partial x_i}$ with $i=x$, $y_1$ and $y_2$ and $\sigma_\xi$ is its root mean square. \textcolor{black}{The probability density function is computed by using $600^2\times 24$ grid points in the mixing layer and $600^2\times120$ grid points in the high energy homogeneous region. In the homogeneous region the probability density function compare well with the data by Ishihara et al. (figure 5 of \cite{is07} and figure 4 of \cite{is09}), interpolated at the same Reynolds number, which use a larger statistical sample ($512^3$).}}
\label{fig.pdf}
\end{figure}

\begin{figure}
{\centering
\includegraphics[width=\columnwidth]{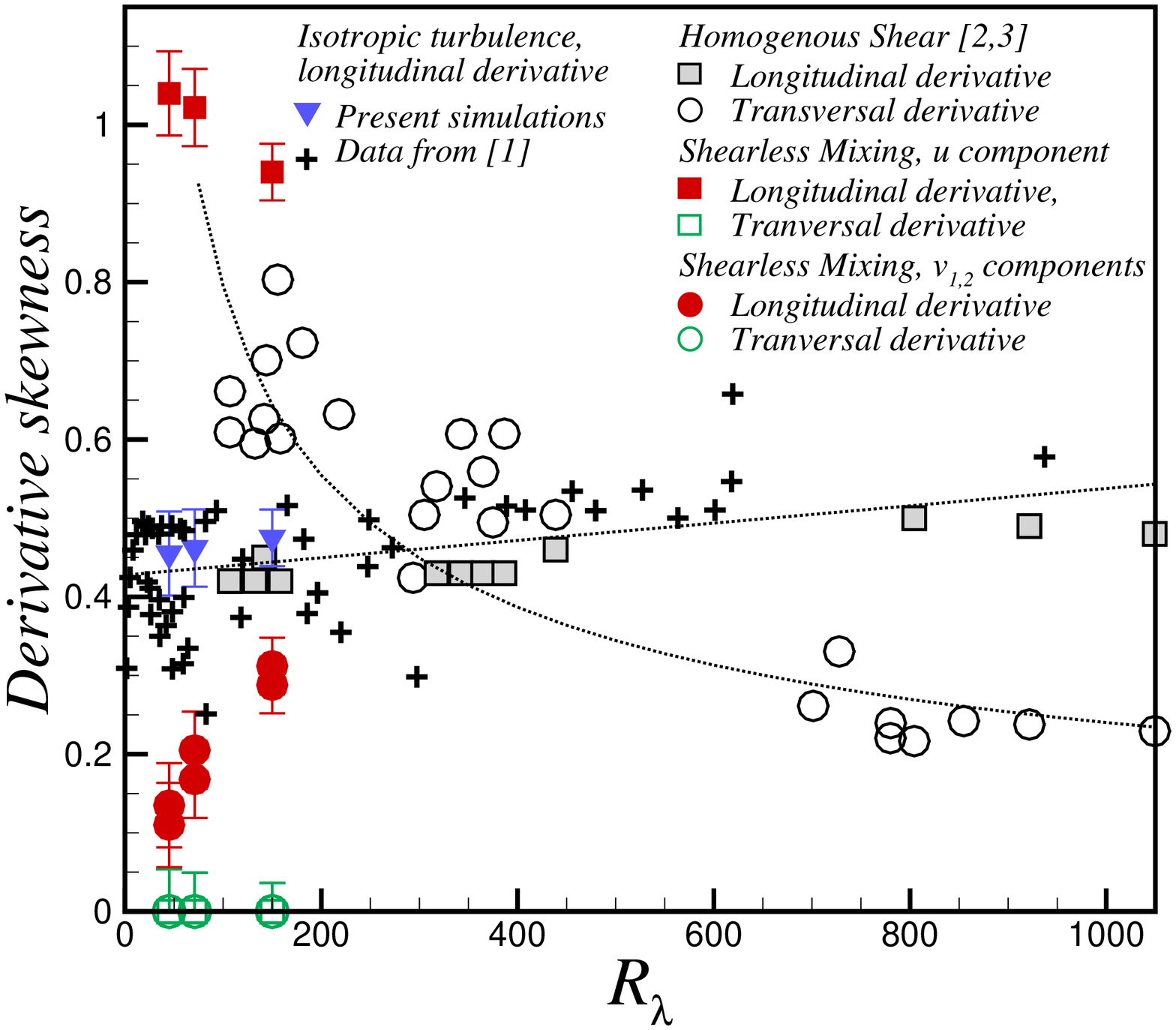} 
}
\caption{Comparison of the derivative skewness modulus in the shearless mixing and in the homogeneous shear flow (data from \cite{ws00} and \cite{ws02}). The data for the homogeneous and isotropic turbulence are taken from the homogeneous regions of present simulations and from \cite{sa97}. Data from ref. \cite{sa97,ws00,ws02} have been read from published graphs.} 
\label{fig.confronto}
\end{figure}

We have analyzed data from numerical simulations in which Navier-Stokes equations have been solved in a parallelepiped domain with a fully dealiased (3/2 - rule) Fou\-rier-Ga\-lerkin pseudospectral spatial discretization and a fourth-order explicit Runge-Kutta time integration \cite{ict01}.
The initial conditions have been obtained from a linear matching of the two homogeneous and isotropic fields over a narrow region - as large as the flow integral scale - by means of a weighting function \cite{ti06,tib08}, see figure 
1 a,b. 
As a consequence, the initial statistics are closer to those present in a homogeneous flow than to those that would emerge across the mixing as the two turbulent flows interact.
A sketch of the computational domain is shown in figure 1 a, $x$ is the coordinate in the inhomogeneous direction and $y_1$, $y_2$ are in the homogeneous directions normal to $x$. The domain has an aspect ratio of 2 (dimensionless size $2\pi^2\times 4\pi$) and the resolution is $128^2\times 256$ at $Re_\lambda=45$, $192^2\times 384$ at $Re_\lambda=71$, $600^2\times1200$ at $Re_\lambda=150$. All the statistical properties of the flow are computed as spatial averages over planes at constant $x$ and are only a function of $(x,t)$.

The two neighbouring turbulent fields are isotropic and each field is statistically defined by the turbulent kinetic energy and the integral scale (or the dissipation rate).
Let us call $E_1$ and $E_2$ the turbulent kinetic energy per unit mass in the two isotropic regions ($E_1>E_2$) and $\ell_1$, $\ell_2$ their integral scales. Three parameters characterize the mixing: the Reynolds number, the  $E_1/E_2$ ratio and the $\ell_1/\ell_2$ ratio.
It has been found that the intermittency level and the depth of penetration by the eddies from the high-energy region increase when the energy and lengthscale gradients are concordant and decrease when they are opposite. Therefore, the most efficient mixing process takes place when the spectra of the two mixed fields differ in the lowest wavenumbers, see \cite{ti06}.

In order to have only one source of anisotropy, the turbulent kinetic energy gradient, two sets of simulations have been considered, both with a uniform integral scale ($\ell_1/\ell_2$=1).
The first set of simulations has an energy ratio fixed at 6.6 and Reynolds numbers, based on the Taylor microscale, of 45, 70 and 150. In the second set, the energy ratio ranges from 6.6 to $10^4$ while the Reynolds number is kept equal to $Re_\lambda=45$.


The discussion focuses on the normalized third and fourth order one-point moments of the longitudinal velocity derivative, that is, on the skewness and kurtosis. These are defined as
\begin{eqnarray}
S_{\partial u / \partial x}&=&\overline{(\partial u / \partial x)^3}/(\overline{(\partial u/\partial x)^2})^{3/2},\nonumber\\
S_{\partial v / \partial y}&=&\overline{(\partial v / \partial y)^3}/(\overline{(\partial v/\partial y)^2})^{3/2},
\label{def.s}
\end{eqnarray}
\begin{eqnarray}
K_{\partial u / \partial x}&=&\overline{(\partial u / \partial x)^4}/(\overline{(\partial u/\partial x)^2})^{2},
\nonumber\\
K_{\partial v / \partial y}&=&\overline{(\partial v / \partial y)^4}/(\overline{(\partial v/\partial y)^2})^{2},
\label{def.k}
\end{eqnarray}
where $y$ is any direction normal to $x$ ($y=y_1$, $y_2$). The overbar denotes the statistical average, which has been approximated by a spatial average on the planes parallel to and inside the mixing layer at a constant $x$, see fig.1 a. The turbulence is homogeneous in these planes.
The velocity fluctuation $u$ in equations (\ref{def.s}, \ref{def.k}) is the component of the velocity vector that is responsible for the energy transport across the mixing.

The spatial distribution of the longitudinal derivative skewness and kurtosis across the mixing layer is shown in figure \ref{fig.sk} at several times, where two mixings with the same energy ratio of 6.6, but a different Reynolds number, are compared. The spatial coordinate $x$ has been rescaled with the mixing layer thickness $\Delta(t)$, conventionally defined as the distance between the points with normalized energy $(E-E_2)/(E_1-E_2)$ equal to $1/4$ and $3/4$ \cite{vw89, ti06,tib08}. Inside the mixing layer, negative values of $\eta=x/\Delta$ correspond to the highest  homogeneous energy flow, positive values to the lowest homogeneous energy  flow.
It can observed that all these statistics depart from the isotropic turbulence value, -0.5, shown in the figure by the horizontal line. However, the longitudinal derivatives exhibit different behavior in different directions.
The main feature of these distributions is that the skewness departs from the opposite sign of the isotropic value: it is negative in the direction normal to the mixing layer (we can observe values as high as -1.1), and is positive in the direction parallel to the layer (values as high as -0.05).

The longitudinal kurtosis shows a maximum in the same zone where the skewness departures are observed, which is always in the low energy side of the mixing. The values of those peaks, for both the skewness and the kurtosis, are almost constant after the initial transient of the simulation, which lasts about 3-4 initial eddy turnover times $\tau=\ell_1/E_1^{1/2}$ when $Re_\lambda=45$, but only about one eddy turnover time when $Re_\lambda=150$;
here the distributions also show a fairly good collapse when the spatial coordinate $x$ is rescaled with the mixing layer thickness $\Delta(t)$. 
A qualitative scheme of the behavior of the longitudinal skewness is shown in figure \ref{fig.schema_long}(a).
With respect to the isotropic situation, the mixing process produces a further compression of the \textcolor{black}{filaments} lying across the mixing layer and a reduction in the \textcolor{black}{filaments} compression in the normal directions. All the mixings follow this common pattern, but the relative values of the deviations from isotropy $\Delta S_{\partial u / \partial x}$ and $\Delta S_{\partial v / \partial y}$ depend on the flow parameters. When the Reynolds number increases, $\Delta S_{\partial v / \partial y}$ 
and $\Delta S_{\partial v / \partial y}/\Delta S_{\partial u / \partial x}
$ decrease (fig.\ref{fig.schema_long}(b)).
An opposite behavior is seen when the energy ratio increases, see again in \ref{fig.schema_long}(b). 
For large values of $E_1/E_2$, the mixing approaches a situation where a turbulent flow diffuses in a region of relatively still fluid, and the main effect on small scales in this limit seems to be an additional negative stretching in the direction of the energy flow.
Figure \ref{fig.pdf} shows the probability density function of the longitudinal derivatives in the mixing, and highlights the longer negative tail of the probability density of the derivative in the kinetic energy flow direction.


A comparison of the longitudinal moments inside the mixing layer with the ones measured in the two homogeneous flows (HIT and homogeneous shear turbulence) is shown in figure \ref{fig.confronto}. The data for the isotropic turbulence in this figure are taken from the homogeneous regions of the present simulations and from the reviews by Sreenivasan and Antonia \cite{sa97}, while the data for the homogeneous shear flows are taken from Warhaft and Shen \cite{ws00,ws02}. These last data are laboratory data which, due to the high level of technical difficulties which characterize this kind of measurements, only give the longitudinal derivative of the velocity component in the streamwise direction. 
The difference with the homogeneous shear flow, which generates large transversal moments, but has less influence on the longitudinal ones (with respect to the isotropic values), is immediately apparent. Shear and shearless flows have thus different kinds of anisotropy in the small scales: a strong differentiation of longitudinal derivative moments for shearless flows and high values of transversal derivative moments for shear flows.

In conclusion, the simulations we have carried out show that a significant small scale anisotropy and intermittency is generated in a decaying shearless turbulent mixing. This intermittency is characterized by a large departure of the longitudinal derivative moments, which are different in the directions across and parallel to the layer, from the typical values of the isotropic condition, even in this flow where there is no energy production (due to the lack of mean flow gradients). The deviations from the isotropic values are large and follow a common trend: the longitudinal derivatives in the energy gradient direction are more intermittent, while the intermittency is milder in the orthogonal directions. 
We also observe that a small intermittency on transversal 
velocity derivatives does not necessarily mean a tendency 
towards isotropy. \textcolor{black}{ The structure of the anisotropy is such 
that the skewness departure from isotropy reduces the 
compression on fluid filaments parallel to the mixing layer 
and enhances that of the filaments orthogonal to it.
In the shear-free mixing, the small-scale turbulence has a different structure than in the homogeneous shear flow case \cite{sa97,ws00,ws02, ssy03, ws04}. The symmetry is different and the anisotropy here is due to the inhomogeneity.  We think that this is the principal agent for the difference in the anisotropy structure of the small-scale in these two flows. 
The reduction of the skewness negativity in directions parallel to the mixing (relative elongation with respect to the isotropic situation) and the enhancement of the negativity across the mixing (relative compression with respect to isotropy) should also be  linked to the incompressibility of the flow.} These effects also persist at moderate Reynolds numbers, therefore the asymptotic approach to the local isotropy, if present, is very slow.



\vspace*{2mm}
\begin{acknowledgments}
\noindent {\bf Acknowledgements:}\\
We are grateful to Z. Warhaft for helpful discussions over the
course of this work. We wish to acknowledge the CINECA supercomputing center support.
\end{acknowledgments}

\end{document}